\documentclass{article}
\usepackage{graphicx}
\usepackage{amsmath,amssymb}
\usepackage{xcolor}
\usepackage{hyperref}
\usepackage{nameref}
\usepackage{titlesec}
\usepackage{authblk}
\title{\textbf{High-Performance Wavelength Division Multiplexers Enabled by Co-Optimized Inverse Design}}

\author[1]{Sydney Mason}
\author[1,*]{Geun Ho Ahn}
\author[1]{Jakob Grzesik}
\author[1]{Sungjun Eun}
\author[1,**]{Jelena Vučković}
\affil[1]{\small{E.L. Ginzton Laboratory, Stanford University, Stanford, CA 94305, USA}}
\affil[*]{\url{gahn@stanford.edu}}
\affil[**]{\url{jela@stanford.edu}}
\date{}
\begin{document}
\maketitle

\begin{abstract}
Wavelength division multiplexers are fundamental to the functioning and performance of integrated photonic circuits, with applications ranging from optical interconnects to sensing and quantum technologies. Current solutions are limited by trade-offs between channel spacing, crosstalk, insertion loss, and device footprint. Here, we develop a novel design approach that co-optimizes inverse-designed wavelength division multiplexers and distributed Bragg gratings to achieve ultra-low crosstalk without compromising insertion loss. We experimentally demonstrate less than -40 dB crosstalk for wavelength channel spacing of 15 nm in foundry-compatible silicon and silicon nitride devices across the telecommunications C- and L-bands. Our design process is highly adaptable, allowing for seamless scaling to a greater number of output channels, different spectral windows, and easy translation across various material platforms. 
\end{abstract}

\bigskip
\section{Introduction}\label{Introduction}
Integrated photonics has emerged as a scalable platform for data communication, becoming prevalent for use in optical interconnects for data centers \cite{DataCenters}. To scale transmission bandwidth, many systems have exploited multi-wavelength data carriers—such as frequency combs or quantum dot mode-locked lasers—to achieve narrowly spaced wavelength channels spanning up to an octave, enabling hundreds of multiplexed information channels \cite{CombLink, dumont2022high, BasicComb, OctaveComb}. Despite the benefits that a dense frequency source offers for optical data communications, the presence of widely spanned narrow channels introduces increased complexity in preserving the signal integrity required for on-chip modulation and data encoding. There have been many approaches to on-chip wavelength de-multiplexing (WDM) including but not limited to arrayed waveguide gratings \cite{AWGsmallish}, thermally tuned ring resonators \cite{RingWDM1, RingWDM2} and inverse design \cite{invdesOG, invdes40nm}. To %[we say "preserve signal integrity twice"]
 preserve signal integrity, it is essential for WDM solutions to exhibit low insertion loss and crosstalk. Arrayed waveguide gratings (AWGs) operate on principles from free space diffraction gratings and have been used in silicon integrated photonic circuits for many decades \cite{AWGOG}. More recent advances in different material platforms have motivated the development of AWGs with materials such as silicon nitride \cite{nitrideAWG} and thin film lithium niobate \cite{tflnAWG} which have low loss and nonlinearity, useful for other optical interconnect components. Although AWGs can achieve sub-3 dB insertion loss \cite{AWGsmallish, tflnAWG} and less than 4 nm channel spacing \cite{tflnAWG}, the requirement for large propagation lengths in both the arrayed waveguides and the free propagation region restricts the miniaturization of AWGs. This results in AWG solutions with device footprints on the order of 100 $\times$ 100 $\mu$m$^2$. Tunable ring resonators can act as filterbanks, using cascaded rings to achieve flat-top responses for pass band channels \cite{RingWDM1, RingWDM2, RingWDM3}. In this approach, the refractive index of the rings are thermo-optically tuned to align the resonances with the input channels. Insertion loss of 1.5 dB, extinction ratio greater than 35 dB and channel spacing of 124 GHz have been demonstrated \cite{RingWDM2}. Although the singular rings have small radii (around 6-12 $\mu$m \cite{RingWDM1,RingWDM2}), this approach requires extra energy for the thermal tuning and two rings per output port, increasing the footprint.

Among existing state-of-the-art solutions for wavelength de-multiplexing is inverse design. Advancing computational platforms and communications systems have motivated higher data rates in optical interconnects and consequently more creative design approaches have materialized. Inverse design allows researchers to explore the entire design space, creating compact and efficient photonic devices with optimized dielectric distributions \cite{InvDesOverview}. Using the adjoint method (which only requires two simulations per design iteration) and accelerated electromagnetic simulations, photonic devices can be optimized for a variety of functionalities \cite{InvDesOverview}. Devices that have been demonstrated include grating couplers \cite{InvDesGC}, cavities \cite{InvDesCav}, filters \cite{invDesFilter}, spatial mode de-multiplexers \cite{KYMDMlink} and wavelength de-multiplexers \cite{invdesOG,invdes40nm}. Despite the irregular geometries that arise from inverse design, the scalability of the resulting devices has been proven through the enforcement of foundry fabrication constraints during optimization \cite{foundryinvdes, InvDesFabCon1, InvDesFabCon2}. Although this approach can outperform classical designs and overcome trade-offs between metrics such as crosstalk, channel spacing, insertion loss, and device area, the best inverse design WDM to date is limited in channel spacing to 20 nm and crosstalk at -26 dB \cite{googleOFC}. A significant reduction in crosstalk has important implications for WDM applications and their system-level utility. Crosstalk introduces clear tradeoffs in optical signal-to-noise ratio (OSNR) for optical communications (Supplement I). Additionally, for quantum and precision applications such as atomic trap arrays or optical clocks, where on and off signal integrity is critical, crosstalk suppression exceeding $>$50 dB is essential. Approaches that utilize high extinction ratio components and exploit photonic band gaps are one of the ways to achieve this magnitude of suppression.

\begin{figure}[!ht]
\centering\includegraphics[width=\linewidth]{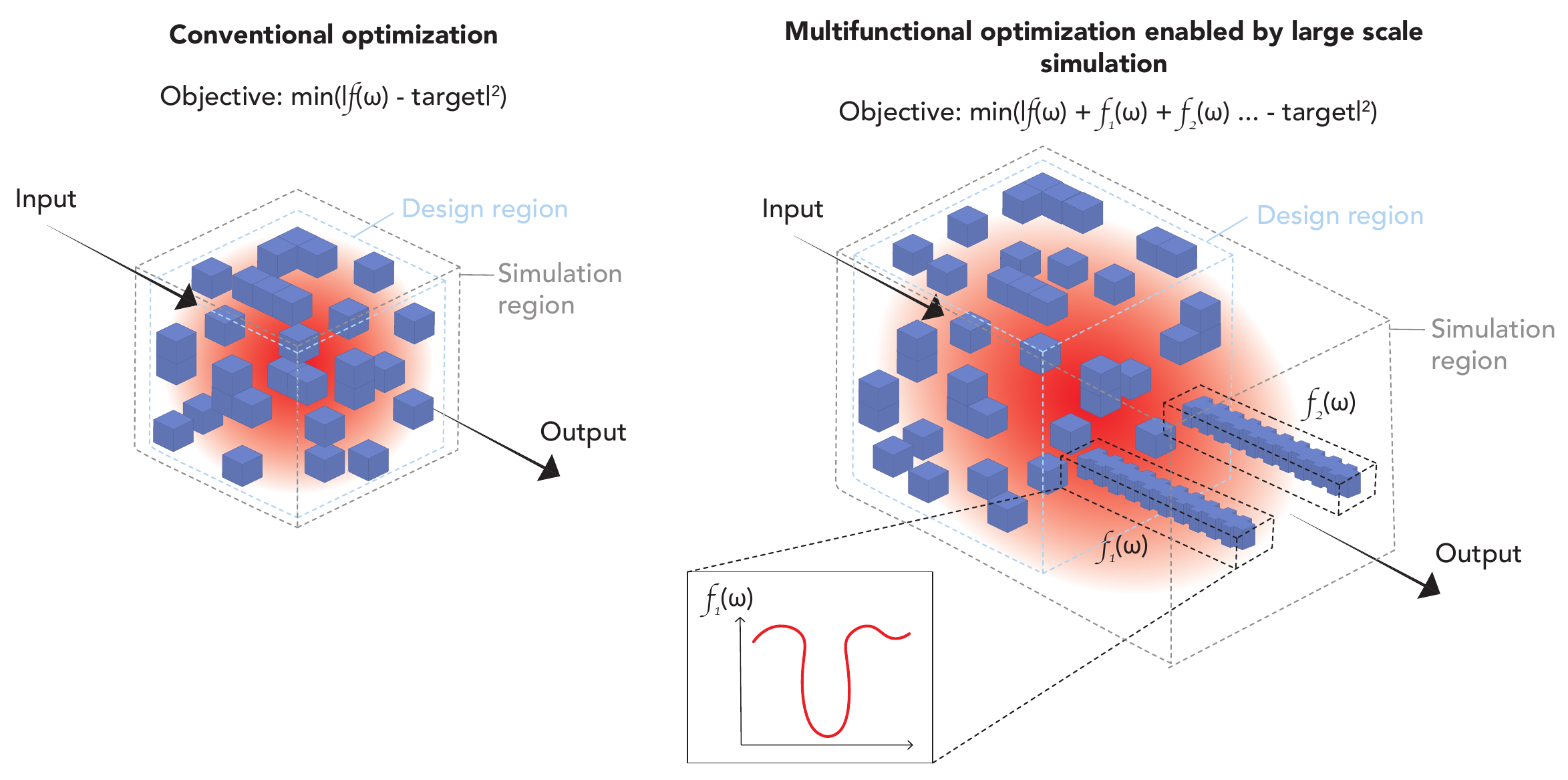}
\caption{\textbf{Approach for next generation of high-performance photonic devices with multifunctional optimization.} Traditional optimization restricts the simulation region to a small design area due to constraints on the computational power of available machines. Now, through the exploitation of GPU architecture and compute speed, we have access to large-scale simulation regions. We demonstrate a design concept for the next generation of multifunctional photonics through the incorporation of existing functionality within the simulation region, resulting in highly performing devices.}
\label{fig:fig1a_conceptualoverview}
\end{figure}

Recent advancements in simulation capacity, specifically GPU-accelerated large-scale FDTD (Finite Difference Time Domain) solvers, have opened the door for the next generation of photonic devices. In free space optics, the design of large-area metasurfaces was previously bottlenecked by simulation volume and computation time. Creative solutions for simulating wide-area metasurfaces \cite{capassolargeareaMS, fanvuckoviclargeareaMS} and GPU-optimized FDTD solvers \cite{fanlargeareaMS, luvuckovicfdtdgpu} have enabled significant progress in free-space optics, which greatly benefits from these larger metasurfaces. In this work, we demonstrate the benefit this recently expanded simulation capacity can have on integrated photonics: multifunctional optimization. Our wavelength de-multiplexers are examples of the highly performing devices that can result from adding a component with some existing functionality to the large simulation region. 

Distributed Bragg waveguide gratings form a one-dimension photonic band gap through periodic perturbations of the refractive index. This band gap makes Bragg gratings excellent candidates for wavelength selective filtering, with near unity transmission of the very narrow pass band (can be as small as 0.5 nm \cite{Bragg0p5nm}). Bragg filters have limited performance capacity in WDM systems to date due to the undesirable, near unity, reflection of the wavelengths in the rejected band. Photonic crystals have been shown to improve the crosstalk of inverse design WDMs when added after optimization is complete \cite{WDMPhC}. In this case, the photonic crystal filters are not included as part of the inverse design process and therefore there is an insertion loss penalty when they are added to the output ports. Moreover, the reflection from the filters that gets re-injected into WDM is neglected, limiting the crosstalk suppression to be -20 dB \cite{WDMPhC}.

Here, we introduce a new approach to the inverse design of WDMs with ultra-low crosstalk through co-optimization with Bragg filters. By initializing the design region with the Bragg filters at the WDM output, the reflection from the Bragg gratings is included in the loss function used for optimization. We achieve narrowband WDM performance by using inverse design to efficiently couple to the Bragg modes. This co-design principle is illustrated in Figure \ref{fig:fig1a_conceptualoverview} and \ref{fig:fig1_devoverview}a. First, we demonstrate this design principle with two-channel silicon WDMs by studying the dependence on Bragg-grating length, showing improved crosstalk suppression without increasing insertion loss. We further show that this method scales to three-channel silicon WDMs. As optical transceivers push data rates beyond 1 Tbps \cite{KYMDMlink, 1tbpsDWDM}, optical power may increase beyond the material limits of silicon \cite{nextGenSi}. Materials like silicon nitride are CMOS foundry-compatible, and offer lower waveguide loss, a lower two-photon absorption coefficient, and greater thermal stability than silicon \cite{SiNvsSi}. However, achieving narrowband WDMs in silicon nitride can be challenging owing to its lower index contrast with oxide cladding as compared with that of silicon. Inverse design has been explored to overcome this constraint with limited cross-talk suppression \cite{invdesSiN}. Therefore, in this work, we employ a co-optimization method for solving SiN wavelength de-multiplexing, highlighting the universality of this design method. We experimentally demonstrate the use of our co-optimization approach to overcome limitations with narrow-band, low crosstalk WDMs in lower index materials and achieve state-of-the-art performance with 15 nm channel spacing, -1.47 dB average insertion loss, and -34.49 dB average crosstalk.

The inclusion of other photonic components as a part of the inverse design process and optimizing the design region in addition to the existing functionality helps to enable the next generation of photonic devices. To further support our proposed co-optimization design methodology, we explore two additional functionalities beyond high-performance WDMs with Bragg filters: (i) a mode converter/wavelength de-multiplexer co-optimized with an inverse design grating coupler, and (ii) a wavelength de-multiplexer co-optimized with inverse design reflectors (Supplement VII). The WDM grating coupler device (i) exhibits -3.47 dB and -6.99 dB of insertion loss and and crosstalk of -29.02 dB and -27.15 dB. The WDM co-optimized with inverse design reflectors (ii) exhibits -1.89 dB of insertion loss for both ports and -29.42 and -34.15 dB of crosstalk for the two ports.% for gc wdm and we achieve -29.42 and -34.15 dB of crosstalk and 1.89 dB for wdm inv des

\section{Results}
\subsection{Wavelength division multiplexer design}\label{Wavelength division mulitplexer design}

\begin{figure}[!htbp]
\centering\includegraphics[width=\linewidth]{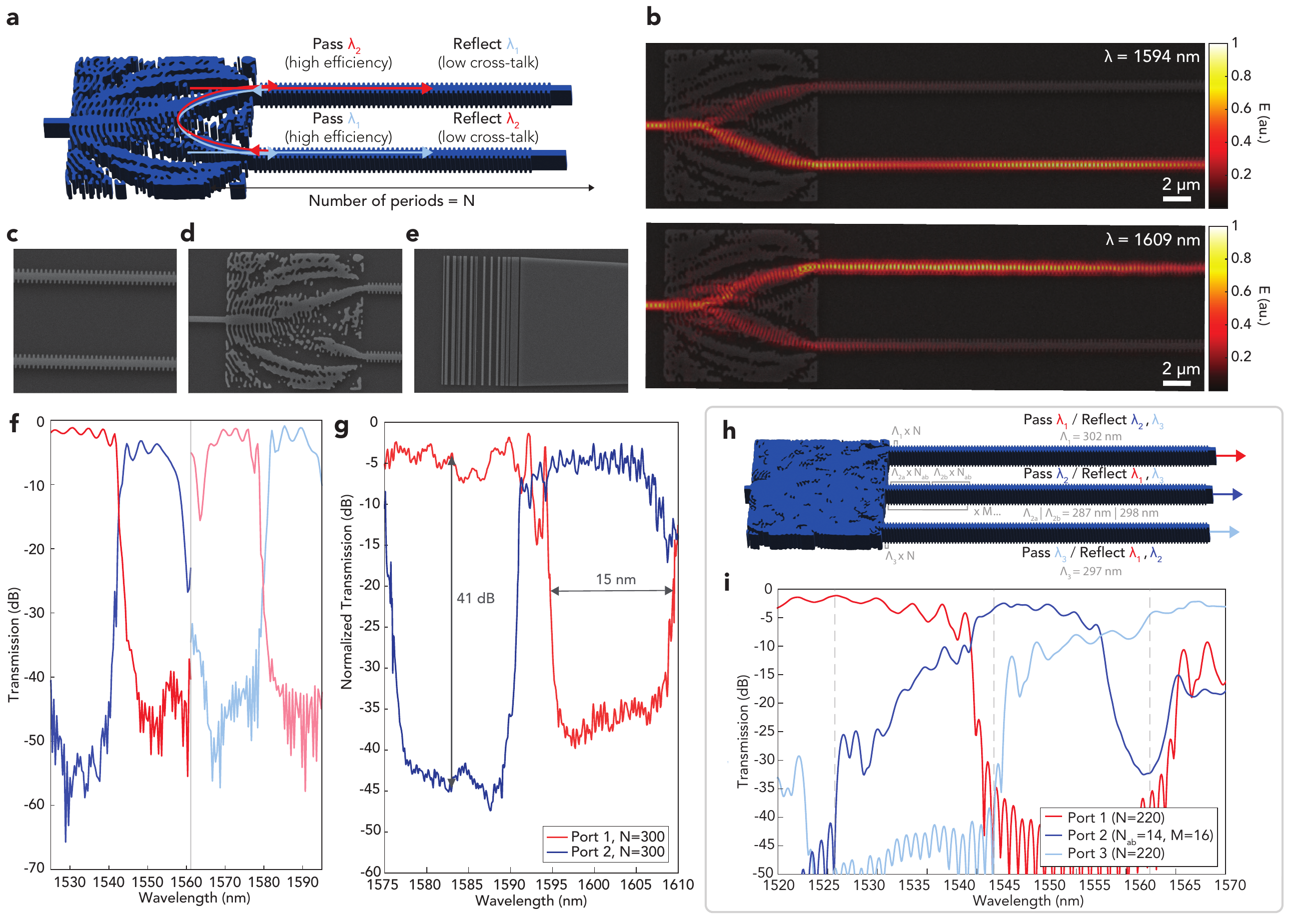}
\caption{\textbf{Design and realization of co-optimized silicon WDMs.} a. Schematic representation of inverse designed Bragg WDMs. Connected to the inverse design region are output ports including Bragg filters that are designed to pass one wavelength channel and reflect the other, b. Scanning electron microscope (SEM) images of the fabricated WDM structures overlayed with the simulated electric field profiles in the structure. Two wavelength channels, 15 nm apart, are shown, c. SEM of the fabricated Bragg filters, d. SEM of one of the WDM devices, e. SEM of the inverse designed grating coupler used to efficiently couple light from free space into the integrated device, f. Simulations of 2 different WDMs designed for different wavelengths. Each of the designs are made with Bragg filters with 200 periods and all simulations are conducted with oxide cladded silicon with a thickness of 220 nm, representative of standard SOI wafers used for fabrication., g. Experimentally measured transmission spectrum of one of the WDM devices. The red and blue lines shows the output power for port 1 and port 2 of the device respectively. h. Schematic representation of a 3-channel WDM co-optimized with Bragg filters. The filters designed to pass $\lambda_1$ and $\lambda_3$ have N periods of a constant periodicity ($\Lambda_{1,3}$) whereas the middle filter has alternating periodicity between two different periods $\Lambda_{2a}$ and $\Lambda_{2b}$. in segments of $N_{ab}$ periods repeated M times., i. Simulated transmission response of the 3-channel silicon WDM. The red line shows the output of port 1, the dark blue shows the output of port 2 and the light blue shows the output of port 3.}
\label{fig:fig1_devoverview}
\end{figure}
Prior to the inverse design of the wavelength de-multiplexers, Bragg gratings are designed to align with the desired WDM output channels. Equations \ref{eq:braggbandwidth} and \ref{eq:braggcentral} govern the bandwidth $\Delta\lambda$ and central wavelength $\lambda_0$ of the Bragg gratings, given the effective refractive indices n$_{eff1}$ and n$_{eff2}$ of the unperturbed and perturbed waveguide sections, the periodicity of the perturbations, $\Lambda$, and the length of the grating $L$. Here, $n_{eff}$ represents the average effective index defined by $n_{eff} = (n_{eff1} + n_{eff2})/2$ and $\kappa$ is the coupling coefficient equal to $\frac{2(n_{eff2} - n_{eff1})}{\lambda_0}$.

\begin{align}
    \Delta\lambda = \frac{\lambda^2_0}{\pi n_g}\sqrt{\kappa^2+\frac{\pi}{L}^2}
\label{eq:braggbandwidth}
\end{align}

\begin{align}
    \lambda_0 = 2\Lambda n_{eff}
    \label{eq:braggcentral}
\end{align}
 
\begin{figure}[!htbp]
\centering\includegraphics[width=\linewidth]{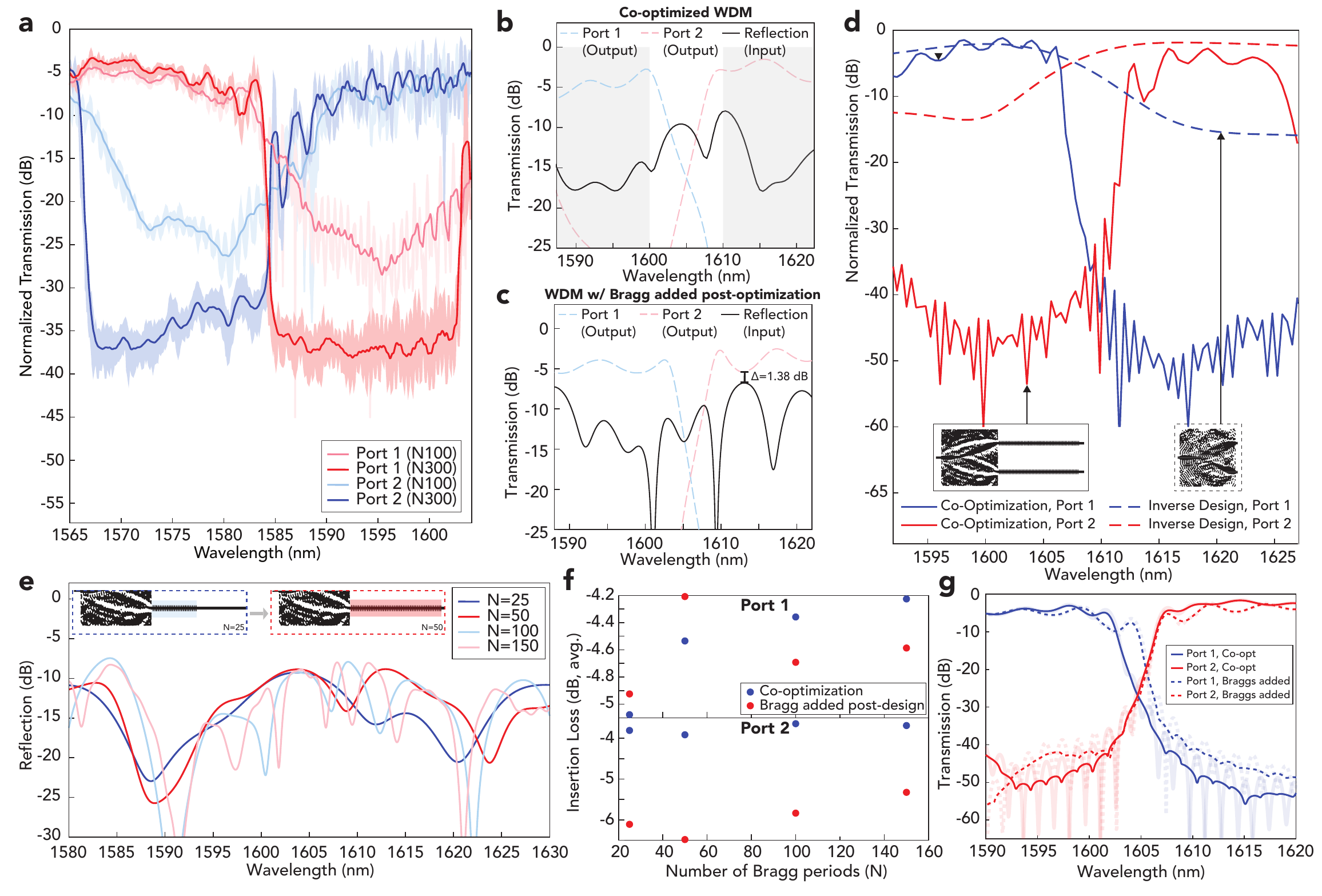}
\caption{\textbf{Comparison of co-optimized devices with traditional inverse design and analysis Bragg filter length dependence.}   a. Experimentally measured difference between devices with different number of Bragg periods at the output. The red lines show port 1 of the same device with differing lengths of Bragg filters (N=100 and N=300) and similarly for port 2 shown in blue. The shaded regions show the minimum and maximum of each port over 5 measurements and the solid lines show the moving average of the 5 measurements, b.,c. Comparison of reflection handling given co-optimized and non-co-optimized case. The device in b. was designed using the larger scale co-optimization which includes the Bragg filters during the inverse design. The gray regions signify the bands for which reflection was minimized during optimization. In c., the device was designed using a basic inverse design process with the Bragg filters added to the structure after the optimization was complete. Here we observe large peaks in the light reflected back into the input,  d. Simulated comparison between WDMs created using traditional inverse design techniques and the technique of co-optimization with Bragg filters used in this work. In the case of co-optimization, there is crosstalk reduction of $>$45 dB and a narrowing of filter bandwidth while insertion loss remains comparable, e. Reflection back into input of WDM device in b., by increasing the Bragg filter length from 25 periods to 150, no significant increase in reflection can be observed, f. Comparison between the insertion loss from co-optimized devices and inverse design WDMs where Bragg filters were added after the fact, g. Comparison between crosstalk of co-optimized devices and inverse design WDMs where Bragg filters were added after the fact. All transmission plots are for Bragg period number N=150 and the shaded regions show the raw data from simulation while the solid and dashed lines show the moving average.}
\label{fig:fig2_reflectionhandling}
\end{figure}

For our first demonstration in a silicon on insulator platform, we design a 2 channel WDM with a channel spacing of 15 nm. Given this channel separation, the two Bragg gratings are designed to have central wavelengths that are 15 nm apart. The bandwidths of the grating are designed such that the edge of one filter aligns with the edge of the other, which gives a bandwidth of 15 nm. The input waveguide width is 500 nm, for single mode operation at telecom wavelengths. The Bragg gratings have widths of 1 $\mu$m with perturbations of 500 nm at a 50\% duty cycle. The period of each Bragg grating varies by device (different wavelength regimes) and material platform (silicon vs. silicon nitride), but, for example for the device whose performance is shown in Figure \ref{fig:fig2_reflectionhandling}d. the Bragg periods are 314 nm and 320 nm to align with output channels at $\approx$1600 nm and 1615 nm. The performance and spectral alignment of the filters are confirmed using FDTD simulations (Spins Photonics \textit{fdtdz}) and imported into the optimization as an initial condition. The design region is defined as a 12 $\mu$m $\times$ 12 $\mu$m region with the two filters at the output ports, separated by 6 $\mu$m. The number of Bragg periods, N, is chosen to be 100 for the inverse design to ensure sufficient reflection of the rejected wavelengths. To guide the optimization, we define a loss function for the gradient descent such that the transmission of each wavelength to the respective output Bragg filter is maximized and the reflection from the Bragg gratings back into the input port is minimized. This objective function is described in Equation \ref{eq:invdeslossfn}. The variable $s_{xyz}$ is used to denote the transmission from the fundamental TE mode of the waveguide of wavelength $x$ from port $y$ to port $z$ where the input waveguide is port labeled port 0 and the two output ports are 1 and 2, the waveguides at the end of the Bragg filters. The coefficients $c_1$ and $c_2$ are used to weight the relative importance of minimizing reflection relative to maximizing the transmission through the device. The flow of optimization is as follows: first, the permittivity distribution of the 2D design region is kept continuous, then after the continuous structure is optimized, the structure is slowly binarized where each pixel (FDTD cell) is either silicon or oxide cladding. Lastly, feature size constraints are imposed in order to have a device that can be feasibly fabricated. We use 100 nm as the minimum feature size for fabrication conducted with university electron-beam lithography tools. This constraint is imposed by slowly ramping the penalization of features smaller than 100 nm in the optimization loss function. For foundry fabrication, this constraint can be tweaked to align with foundry specific DRCs (design rule checks) for minimum feature and gap sizes.

\begin{align}
    f = (|s_{101}|^2 + |s_{202}|^2) - (c_1|s_{100}|^2 + c_2|s_{200}|^2)
    \label{eq:invdeslossfn}
\end{align}

For proof of concept, we design several 2-channel silicon WDM filters across the C- and L-bands of telecommunications, and their simulated transmission spectra are shown in Figure \ref{fig:fig1_devoverview}. This entire inverse design process is conducted using the Spins Photonics \textit{fdtdz} software package.

For the 3-channel design displayed in Figure \ref{fig:fig1_devoverview}h., the Bragg filters require a slightly more careful design. In this case, each Bragg filter now needs to reflect two channels. For the two filters designed to pass the highest and lowest wavelength channels ($\lambda_1, \lambda_3$), we can simply design Bragg filters with twice the bandwidth of the 2-channel case, centered halfway between the central wavelengths of the desired two reflected channels. For the filter designed to pass the middle wavelength channel $\lambda_2$ and reflect $\lambda_1$ and $\lambda_3$, it is necessary to design a Bragg filter with an alternating periodicity. Specifically, for our 3-channel WDM designed for 15 nm channel spacing we design Bragg filters for $\lambda_1$=1530 nm, $\lambda_2$=1545 nm and $\lambda_3$=1560 nm. This results in Bragg periodicities of 302 nm and 297 nm for channel 1 and channel 3, respectively. For the middle channel, alternating periods of 287 nm and 298 nm in sections of 14 periods, repeated four times, yields a suitable reflection profile. The resulting transmission spectra for these filter designs are shown in Supplementary Figure 1b. The design region for the 3-channel case is 14 $\mu$m $\times$ 14 $\mu$m and the three filter waveguides are separated by 5.6 $\mu$m.

The silicon nitride devices are designed using the same process as described previously for the silicon 2-channel devices. To ease the optimization process in a more difficult design space, the design region is expanded from 12 $\mu$m $\times$ 12 $\mu$m to an area of 20 $\mu$m $\times$ 14 $\mu$m. As well, the Bragg filters have different dimensions to account for the shift in mode index and consequently, waveguide dimensions. In this case, the base waveguide for Bragg filter is 1 $\mu$m wide with teeth that are 1.5 $\mu$m in width and the waveguides are separated by 10 $\mu$m at the output. Using Equation \ref{eq:braggbandwidth} and \ref{eq:braggcentral}, we design Bragg filters for a stack with 310 nm thick silicon nitride, air cladded, on a silicon dioxide on silicon substrate. We co-optimize two silicon nitride WDMs with 15 nm channel separation in the C-band. The first WDM has channel wavelengths of 1515 nm and 1530 nm corresponding to Bragg periodicities of 490 nm and 500 nm and the second for 1555 nm and 1570 nm with Bragg periodicities of 506 nm and 512 nm. An overview of the devices as well as experimental performance is shown in Figure \ref{fig:fig5_SiN}.

\subsection{Wavelength division multiplexer performance}
To analyze the benefit of the co-optimized inverse design approach, we examine the performance of inverse designed WDMs when the Bragg filters are not included in the optimization. The co-optimization case is compared with the traditional inverse design approach as well as when the Bragg filters are added to the structure after optimization. This analysis is detailed in Figure  \ref{fig:fig2_reflectionhandling}. A significant reduction in crosstalk and narrowing of bandwidth is observed for the co-optimized WDM and Bragg filter design as compared with simple inverse design (Figure \ref{fig:fig2_reflectionhandling}d.). By including the Bragg filters in the large scale simulation region used for optimization, the resulting structure reduces the reflection back into the input port while also efficiently routing the wavelengths to their respective outputs (difference shown in Figure \ref{fig:fig2_reflectionhandling} b.,c., and e.). 
% Without the co-optimization technique, we observed peaks in the reflection spectrum in the gap between the two filters. Overall we see a reduction in reflection by $\sim$4 dB using the co-optimization method and in one case, we can optimize the structure to reduce the peak in the gap between the filters to around -24 dB.                               
A true advantage to using Bragg filters is the ability semi-arbitrarily reduce crosstalk by simply increasing the number of Bragg periods. This can be done at no expense to insertion loss. This effect is explored in Figure \ref{fig:fig2_reflectionhandling}a. where we measure the same devices but with increasing Bragg filter lengths (indicated by the number of Bragg periods N). By increasing N to 300, crosstalk of less than -40 dB can be achieved while maintaining insertion loss. This approach has a minimal impact on the overall device footprint, as it involves merely extending the already narrow output waveguides. 

The 3-channel silicon device yielded similar performance metrics as compared with the 2-channel device, demonstrating the robustness of our approach to handling the interference and reflection from multiple closely spaced channels. Such handling would be significantly worse for non-co-optimized case. The 3-channel device has a simulated mean insertion loss of -2.83 dB and mean crosstalk of -43.26 dB across the three output ports. We note a small discrepancy between the optimal peak of designed and simulated wavelength channels which may be due to the difference in sub-pixel averaging across FDTD simulators (\textit{fdtdz} and Lumerical).

\begin{figure}[!ht]
\centering\includegraphics[width=\linewidth]{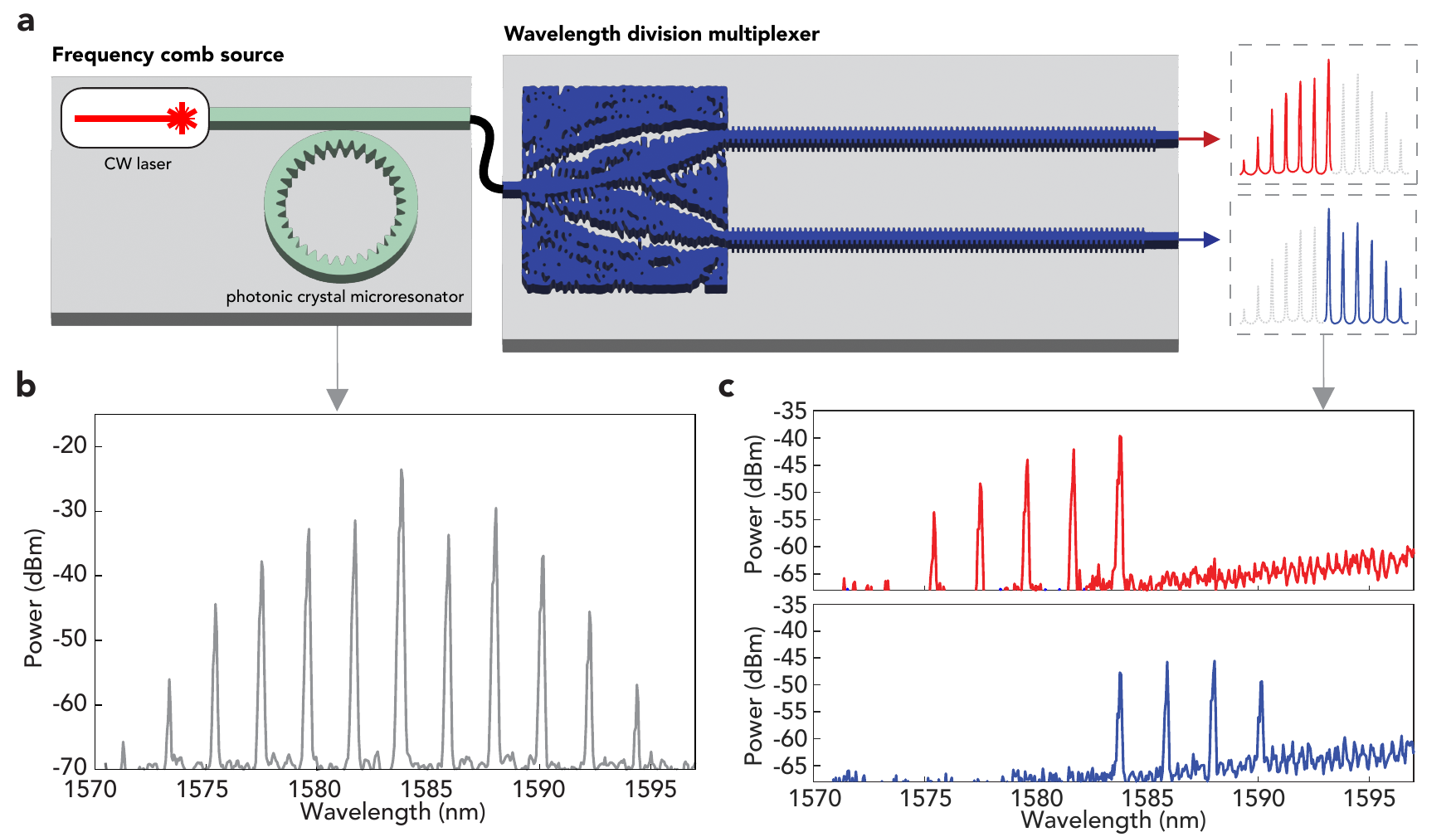}
\caption{\textbf{Experimental demonstration of WDM devices.} a. Schematic representation of system demonstration of the WDM device coupled to a frequency comb source for filtering of the comb lines.  A frequency comb is generated from a photonic crystal microresonator in silicon nitride and this comb is sent to the WDM chip. The resulting outputs of the two ports are measured using a photodiode and power meter, b. Measured input comb spectrum from an OSA, c. Measured output spectra from each of the two WDM ports, normalized to the response of grating couplers used to couple light in and out of the WDM chip.}
\label{fig:fig4_comb}
\end{figure}

Additionally, we explore an application of the WDM devices at a system level. We examine the integration of our WDM solution with a multi-channel source by coupling one of the WDM devices with a silicon nitride frequency comb source to separate the comb lines into each of the WDM output ports. The frequency comb source is generated using a photonic crystal microring resonator which gives the input spectrum shown in Figure \ref{fig:fig4_comb}b.. The resulting output spectra from the two ports of the WDM are shown in Figure \ref{fig:fig4_comb}c., where we normalize to the response of input and output grating couplers. Four comb lines are coupled into the output of one port of the WDM while three comb lines are coupled into the other port. As observed in Figure \ref{fig:fig4_comb}c., the pump for the frequency comb is coupled to both of the WDM outputs due to the higher power relative to other comb lines.

\begin{figure}[!ht]
\centering\includegraphics[width=\linewidth]{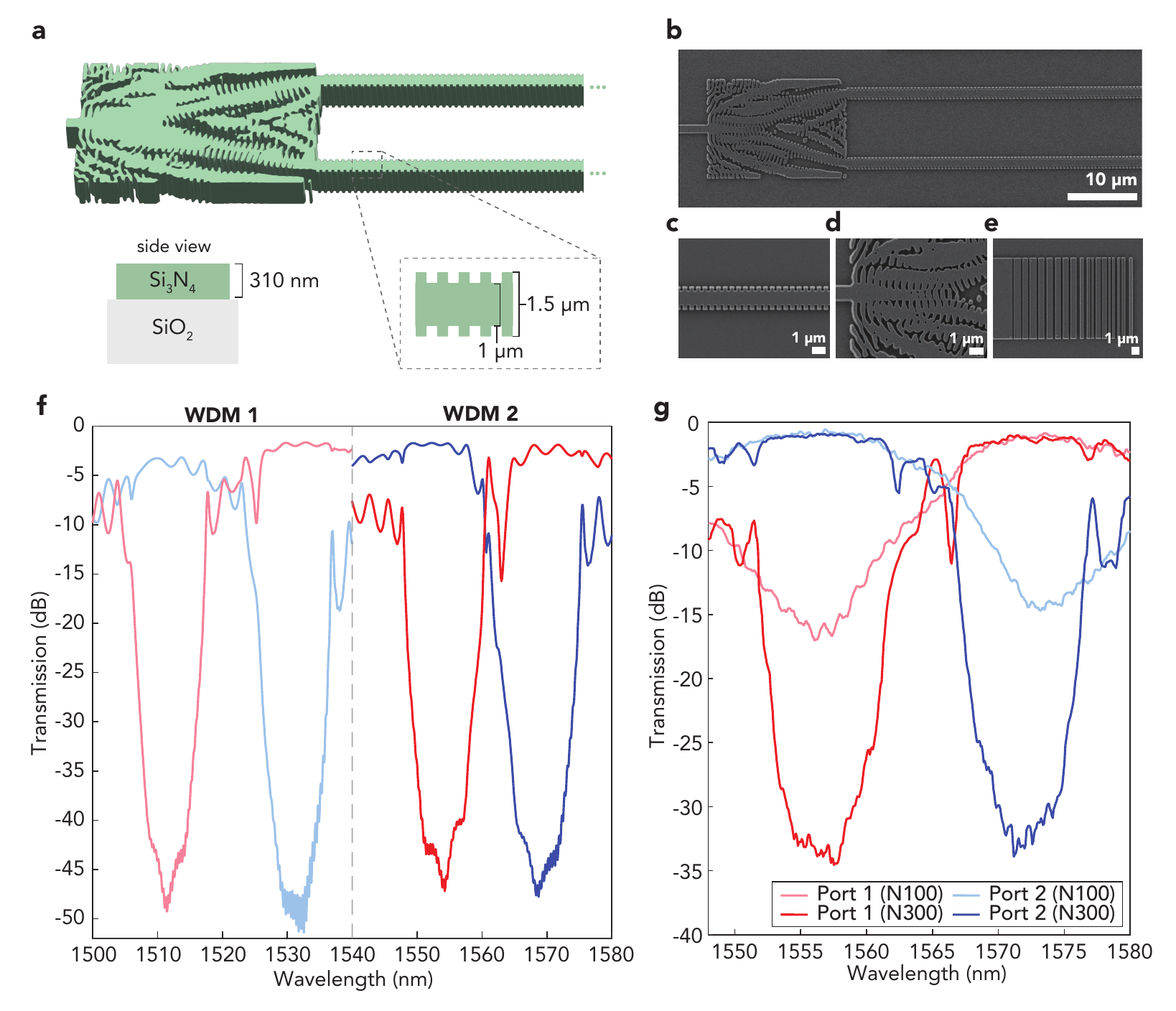}
\caption{\textbf{Silicon nitride WDM design and performance.} a. Schematic representation showing the resulting design for a co-optimized WDM in silicon nitride. The fabricated material stack consists of air-cladded 310 nm thick silicon nitride on SiO$_2$. The inset shows the Bragg filter dimensions., b. Scanning electron microscope (SEM) image of one of the fabricated silicon nitride structures., c. SEM of Bragg waveguide filter., d. SEM zooming in on inverse design region, showing fidelity of small fabricated features., e. SEM of 1D inverse design grating coupler used to efficiency couple light from free space into the fundamental waveguide mode., f. Simulated transmission efficiency of two silicon nitride WDMs (number of Bragg periods N=300) where less than -45 dB of crosstalk can be observed., g. Experimental measurement of WDM 2 from f.. The light red and light blue lines show the transmission spectra of devices with N=100 and the dark red and dark blue shows devices with N=300. An increase in crosstalk suppression while maintaining low insertion loss can be observed with the increase in N.}
\label{fig:fig5_SiN}
\end{figure}

To exemplify the universality of this design method, we fabricate and experimentally demonstrate a silicon nitride WDM. The silicon nitride is deposited using low pressure chemical vapor deposition (LPCVD), yielding a low-loss stoichiometric silicon nitride, the same material that is used for the frequency comb generation in Figure \ref{fig:fig4_comb}. Thus, this would allow for easy integration with a frequency comb source on the same chip in future devices which could significantly reduce power loss due to coupling on and off chip between the comb source and WDM. The fabrication process is further described below in \nameref{Methods}. As with the silicon devices, we analyze the silicon nitride device performance with a varying Bragg period number N, observing an increase in crosstalk suppression and low constant insertion loss with the longest Bragg filter length (Figure \ref{fig:fig5_SiN}f.). For a WDM device designed for 1510 and 1530 nm, the optimal center channel wavelengths in simulation are 1514 and 1530 nm, exhibiting insertion loss of -3.15 dB and -1.70 dB and crosstalk of -47.79 dB and -48.20 dB respectively. For a second WDM device designed for longer wavelengths, the insertion loss at the optimal channel wavelengths (slightly shifted from the design, ~2 nm for second channel) of 1555 nm and 1568 nm is -1.75 dB and -1.91 dB with crosstalk of -45.65 dB and -46.06 dB respectively. In Figure \ref{fig:fig5_SiN}f., we show the experimental data for the WDM designed for 1550 nm/1570 nm channels since this device aligns best with the laser used for measurement (1510 - 1630 nm range). We measure crosstalk values of -34.50 dB and -34.48 dB and average insertion loss values of -1.325 dB and -1.616 dB for the devices with Bragg period number N=300. To the best of our knowledge, these values are the lowest crosstalk value reported in an inverse design silicon nitride WDM to date. We attribute this difference in insertion loss between the simulated and experimentally measured devices to the fabrication process (under-etching and small refractive index variations). In our devices, we note that there can be a difference in insertion loss between the two ports of a single WDM. Although this difference can be observed in the transmission plots, an analysis of the average insertion loss difference shows the average difference between two ports of a single device to be a reasonable $\Delta_{loss}=$ 1.21 dB. This average is calculated across all of the WDM performances shown in this work, both experimental and simulated and silicon and silicon nitride devices and is calculated for a 10 nm band centered around the Bragg filter center wavelength.

\section{Discussion}\label{Discussion}
Through co-optimization of distributed Bragg gratings and dielectric wavelength division multiplexers, we achieve low insertion loss and narrowband operation with ultra-low crosstalk. This design process is compatible with foundry fabrication protocols and easily transferable to different materials and material stacks, as demonstrated by the devices in both silicon and silicon nitride. More broadly, this framework suggests a pathway toward scalable photonic-circuit synthesis in which circuit subsystems are treated as composite scattering elements (i.e., effective multiport $S$-matrices) whose responses can be optimized under realistic coupling and reflection environments. In this paradigm, individual components can be co-designed with their surrounding context, and subsystem performance can be systematically improved by optimizing either the composite structure or selected elements within it, informed by full-system electromagnetic simulations. This capability is increasingly relevant as photonic circuits grow in complexity and density, where accumulated parasitic interactions increasingly limit performance. Consequently, co-optimized inverse design provides a general and accurate route to large-scale photonic-system optimization and automation, innovating the high performance photonic systems.

\bigskip

\section{Methods}\label{Methods} 
\subsection{Device Fabrication}
The silicon devices are fabricated on a silicon on insulator (SOI) substrate with a silicon device layer of 220 nm. The inverse-designed WDM and Bragg structures are patterned using electron beam lithography and positive resist (ZEP520A). The resist is developed in solvents (xylene then methyl isobutyl ketone (MIBK) and isopropyl alcohol (IPA)) and the silicon is etched using reactive ion etching (RIE). A HBr and Cl$_2$ chemistry is used for the RIE. The resist is removed (using Remover 1165) and the substrate is cleaned in piranha solution (sulfuric acid and hydrogen peroxide) and an HF dip prior to oxide cladding. The silicon structures are cladded in oxide through spin coating of hydrogen silsesquioxane (HSQ) and annealing in nitrogen gas at 900$^\circ$C.

The silicon nitride devices are fabricated on a LPCVD silicon nitride layer that is deposited on a silicon substrate with 5 $\mu$m SiO$_2$ layer on top. The silicon nitride is deposited at 830$^\circ$C furnace yielding low-loss stoichiometric silicon nitride. As with the silicon devices, the WDMs are patterned and developed using the same electron beam lithography process described above. The silicon nitride is etched using reactive ion etching with a CF4 and CHF3 chemistry. The resist is removed (using Remover 1165) and the substrate is cleaned in piranha solution (sulfuric acid and hydrogen peroxide) and an HF dip.

\subsection{Experimental Setup}
The transmission characteristics of the devices are measured experimentally using a fiber-to-chip grating coupler setup. Input light from a telecom laser (Toptica, 1510-1630 nm) is swept continuously while the output power from the WDM ports is measured simultaneously on an optical power meter (Newport 1919-R). The transmission of the devices is normalized to the output power spectrum of the two grating couplers, to abstract the performance of the WDM device alone.

\section{Data Availability}

The datasets generated during and/or analysed during the current study are available from the corresponding author on reasonable request.

\section{Code Availability}

The code generated for the current study are available from the corresponding author on reasonable request.

\section{Acknowledgements}\label{Acknowledgements}
\subsection{Funding}
This work is funded by DARPA under the PIPES program and AMD through the System-X Alliance at Stanford. 
J.G. acknowledges support from the Hertz Foundation Graduate Fellowship. S.E. acknowledges support from the Shoucheng Zhang Graduate Fellowship and the Korea Foundation for Advanced Studies Overseas Ph.D. Fellowship.
Part of this work was performed at the Stanford Nanofabrication Facility (SNF) and the Stanford Nano Shared Facilities (SNSF), supported by the National Science Foundation under Grant No. ECCS-2026822.

\subsection{Author Contributions}
S.M., G.A., and J.V. conceived of the project. S.M. designed, simulated, and measured the devices. S.M., J.G., and S.E. fabricated the devices. J.G. designed and fabricated the frequency comb source, and J.G. and S.M. conducted the frequency comb measurements.  G.A. and J.V. supervised the project. All authors contributed to the writing of the manuscript and the analysis of the results.

\subsection{Competing Interests}

The authors declare the following competing interest:

Patent Application No. PCT/US2025/036120.

\end{document}